\documentclass{aastex6}

\fullcollaborationName{The Friends of AASTeX Collaboration}

\begin{document}


\title{Characterization of the VVV Survey RR Lyrae Population across the Southern Galactic Plane}


\author{Dante Minniti\altaffilmark{1,2,3} \and Istvan D\'ek\'any\altaffilmark{4} \and Daniel Majaess\altaffilmark{5,6} \and Tali Palma\altaffilmark{2,1} \and Joyce Pullen\altaffilmark{2,1} \and Marina Rejkuba\altaffilmark{7,8} \and Javier Alonso-Garc\'ia\altaffilmark{9,2} \and Marcio Catelan\altaffilmark{10,2}, \and Rodrigo Contreras Ramos\altaffilmark{2,10}  \and Oscar A. Gonzalez\altaffilmark{11}  \and Maren Hempel\altaffilmark{10} \and Mike Irwin\altaffilmark{12} \and  Philip W. Lucas\altaffilmark{13}  \and Roberto K. Saito\altaffilmark{14} \and Patricia Tissera\altaffilmark{1,2} \and Elena Valenti\altaffilmark{7} \and Manuela Zoccali\altaffilmark{10,2}}

\altaffiltext{1}{Departamento de Ciencias F\'isicas, Facultad de Ciencias Exactas, Universidad Andr\'es Bello, Av. Fern\'andez Concha 700, Las Condes, Santiago, Chile.}
\altaffiltext{2}{Instituto Milenio de Astrof\'isica, Santiago, Chile.}
\altaffiltext{3}{Vatican Observatory, V00120 Vatican City State, Italy.}
\altaffiltext{4}{Astronomisches Rechen-Institut, Zentrum fuer Astronomie der Universitaet Heidelberg, Moenchhofstr. 12-14, D-69120 Heidelberg, Germany.}
\altaffiltext{5}{Mount Saint Vincent University, Halifax, Nova Scotia, Canada.}
\altaffiltext{6}{Saint Mary's University, Halifax, Nova Scotia, Canada.}
\altaffiltext{7}{European Southern Observatory, Karl-Schwarszchild-Str. 2, D85748 Garching bei Muenchen, Germany.}
\altaffiltext{8}{Excellence Cluster Universe, Boltzmannstr. 2, 85748, Garching, Germany}
\altaffiltext{9}{Unidad de Astronomia, Facultad Cs. Basicas, Universidad de Antofagasta, Avda. U. de Antofagasta 02800, Antofagasta, Chile.}
\altaffiltext{10}{Pontificia Universidad Cat\'olica de Chile, Instituto de Astrofisica, Av. Vicuna Mackenna 4860, Santiago, Chile.}
\altaffiltext{11}{Institute for Astronomy, University of Edinburgh, Royal Observatory, Blackford Hill, Edinburgh, EH9 3HJ, UK.}
\altaffiltext{12}{Institute of Astronomy, Cambridge University, Cambridge, CB3 0HA, UK.}
\altaffiltext{13}{Department of Astronomy, University of Hertfordshire, Hertfordshire, UK.}
\altaffiltext{14}{Departamento de F\'{i}sica, Universidade Federal de Santa Catarina, Trindade 88040-900, Florian\'opolis, SC, Brazil.}


\begin{abstract}
Deep near-IR images from the VISTA Variables in the V\'ia L\'actea (VVV) Survey were used to search for RR Lyrae stars in the Southern Galactic plane.  A sizable sample of 404 RR\,Lyrae of type ab stars were identified across a thin slice of the 4$^{\rm th}$ Galactic quadrant ($295\degr < \ell < 350\degr$, $-2.24\degr < b < -1.05\degr$).  The sample's distance distribution exhibits a maximum density that occurs at the bulge tangent point, which implies that this primarily Oosterhoff type I population of RRab stars does not trace the bar delineated by their red clump counterparts. The bulge RR\,Lyrae population does not extend beyond $\ell \sim340 \degr$, and the sample's spatial distribution presents evidence of density enhancements and substructure that warrants further investigation. Indeed, the sample may be employed to evaluate Galactic evolution models, and is particularly lucrative since half of the discovered RR\,Lyrae are within reach of Gaia astrometric observations.  
\end{abstract}




\section{Introduction} \label{sec:intro}
Optical surveys (e.g.\ OGLE, La Silla Quest, LINEAR, Catalina, and other Surveys) have identified thousands of RR\,Lyrae stars, which are key tracers of old and metal-poor regions of the Galaxy.  In particular, RR\,Lyrae (1) are excellent reddening indicators, as the stars span a narrow range of intrinsic colors; (2) are primary standard candles that adhere to well-defined period-luminosity relations; and (3) the pulsators are proxies for age and metallicity, as they constitute an old (age $ >10$\,Gyr) and metal-poor demographic linked to the primordial stellar population of the Milky Way.  This class of variable stars was used to characterize various components of the Milky Way, including the halo \citep[][etc.]{miceli, drake, vivas, helmi08, helmi11}, the bulge \citep[][etc.]{alcock, kunder08, dekany, pietrukowicz15, gran16}, the Solar neighbourhood \citep{layden94, layden98}, and the outer thick disk towards the anticenter \citep{sesar}.   
RR Lyrae  could be used as tracers of accreted galaxies  which might have contributed to the formation of the inner stellar halo-bulge transition region as suggested by hydrodynamical simulations of MW mass-like galaxies (Tissera et al. 2017).

In a pioneering work \citet{layden94} presented a catalog of RR\,Lyrae within 5.5\,kpc of the Sun, and the sample subsequently became a suite of well-studied local calibrators. Yet, it is challenging to identify RR\,Lyrae throughout the complex Galactic plane region beyond the Solar neighborhood and toward the inner disk, where sizeable extinction can render the apparent magnitude beyond the faint limit of most optical surveys.\\

VISTA Variables in Via L\'actea (VVV), a deep near-infrared (NIR) survey, was consequently carried out to identify and characterize RR\,Lyrae stars in some of the most crowded and reddened regions of the Galaxy.  The survey sampled 562\,sq. deg. of the bulge and the southern disk \citep{minniti10, saito, hempel} enabling discovery of numerous classes of variable stars concurrently with the RR Lyrae variables, such as bright LPVs/Miras, eclipsing binaries, Cepheids, and microlensing events \citep{catelan13,minniti15}. RR\,Lyrae stars are the focus of the present study, which are concentrated on the lower latitude of the southern disk and its substructures.  \\

The present search for new RR\,Lyrae candidates expands in part upon our previous work.  \citet{dekany} studied bulge RR\,Lyrae by combining optical with NIR photometry, and discovered that the inner bulge RR\,Lyrae do not display the barred structure traced by red clump giants, but rather trace an inner spheroidal component.   \citet{gran16} searched for new RR\,Lyrae in the bulge-halo transition region, and identified 1019 new RR\,Lyrae type ab variables (RRab)  measuring their parameters that included reddening and distance. Those outer bulge RR\,Lyrae likewise did not adhere to a barred structure.  The present work describes the results based on an extended search for RRab stars at low latitudes, whose broader objective is to map the outer bar region and the inner disk. We present the discovery of 404 new RRab stars, which occupy a low latitude slice across the Galactic plane.\\

This paper is organized as follows.  Section~\ref{sec:selection} describes the selection of the RR\,Lyrae stars, while Section~\ref{sec:CMDs} presents color-magnitude diagrams (CMDs) for our sample. Section~\ref{sec:reddening} discusses the reddening and extinction estimates established, whereas the computed distances are provided in Section~\ref{sec:distance}, which likewise includes a discussion concerning the sample's spatial distribution. In Section~\ref{sec:metallicity} we discuss the inferred metallicity estimates, and Section~\ref{sec:populations} presents a comparison with other RR\,Lyrae populations. Finally, the conclusions are summarized in Section~\ref{sec:conclusion}. \\

\section{The VVV Survey RR Lyrae Selection} \label{sec:selection}

\begin{figure}[!th]
\centering
\includegraphics[width=0.8\hsize]{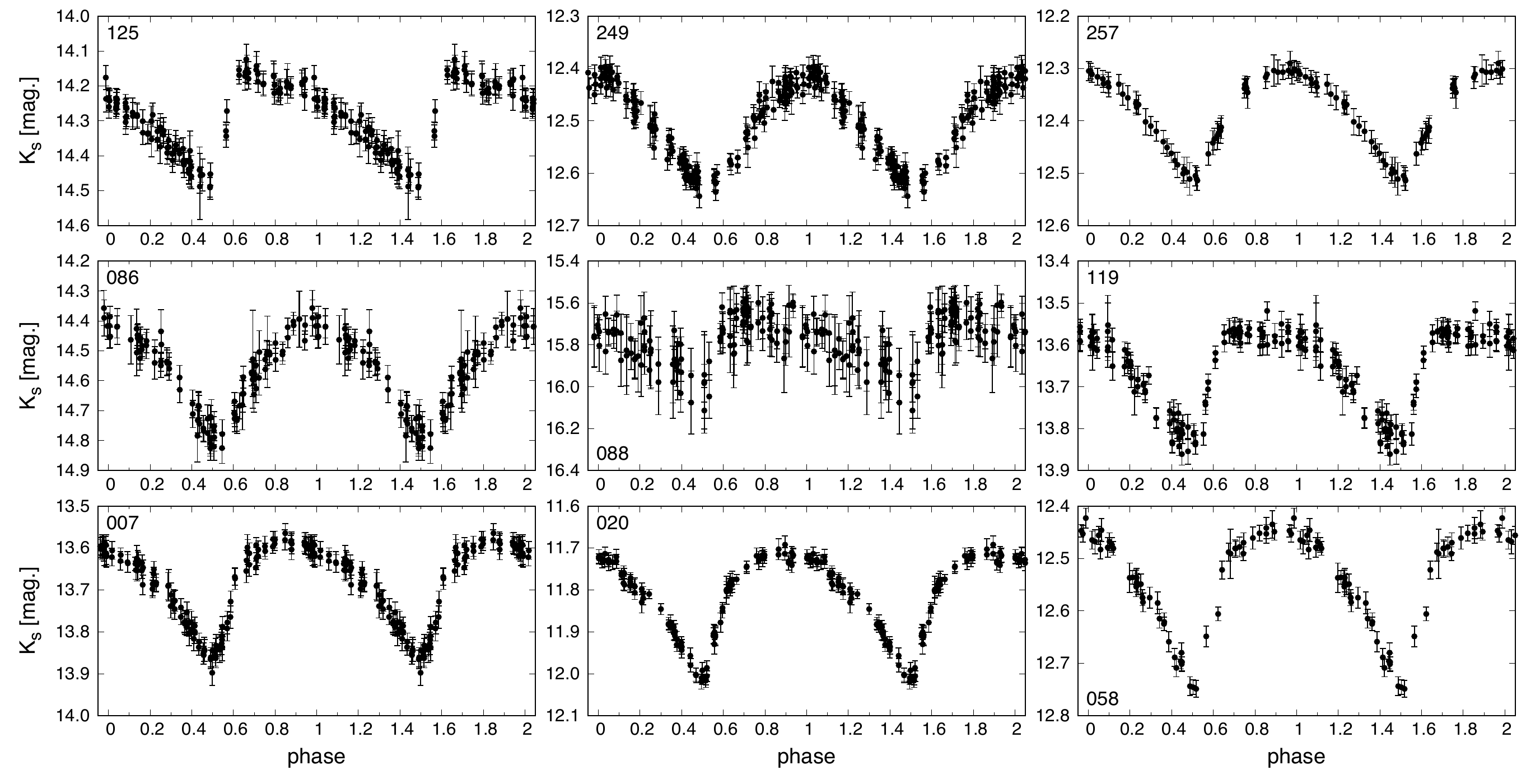}
\caption{Sample phased light curves for some candidate RR\,Lyrae type ab. For this work we have concentrated on the search for RRab stars, that are easier to classify because they have asymmetric light curves. \label{f:fig1}}
\end{figure}

The NIR VVV Survey observations \citep{minniti10, saito, hempel} were  acquired with the VIRCAM camera at the VISTA 4.1\,m telescope at ESO Paranal Observatory (Emerson \& Sutherland 2010). In the disk fields typically 70 epochs of observations were acquired in the $Ks$-band between years 2010 and 2015, plus complementary single epoch observations in the  $ZYJH$ bands. The 16 NIR detectors of VIRCAM produce an image of 11.6'$\times$11.6'  and a pixel scale of 0.34''/pixel. The deep multi-epoch $K_s$ band photometry allows us to unveil faint variable sources deep in the disk regions of our Galaxy (Figure~\ref{f:fig1}). A search for RRab stars was made throughout tiles d001 to d038 of the VVV survey's disk field, which is a thin slice through the Galactic plane spanning $295 < \ell < 350 \degr$, and $-2.24\degr < b < -1.05\degr$.  The total area covered is nearly 57\,sq.deg, and the limiting magnitude in the specified tiles spans 17.5 to 18.5 mag in the Ks-band \citep[see the tile positions and maps of][]{saito}.\\

The search concentrated on RRab stars, because these objects are readily identifiable via their asymmetric light curves, which feature a saw-tooth morphology \citep{layden97}.  Contrarily, RRc variables feature more sinusoidal light curves, and increased contamination can occur from numerous short period eclipsing binaries that have similar light curve morphology. As a matter of fact, RRc type are about $70\%$ as numerous as RRab type stars in an Oosterhoff I population, and future searches must consider additional properties to select \textit{bona fide} samples of the former.  Amplitudes of NIR RR\,Lyrae light curves are one third of those in the optical \citep[e.g][]{navarrete}. \\

The procedures outlined by \citet{gran15} and \citet{javier} were adopted to identify RRab variables.  Approximately $\sim$130,000 variable star light curves were inspected, and the final selection required for periodic objects to fulfill the following criteria: periods spanning $0.38\leqslant P \leqslant1.0$ days, displaying amplitudes of $A_{K_s} \geqslant 0.1$ mag, and featuring mean magnitudes of $11.5 \leqslant Ks \leqslant 16.75$.  A limit on the maximum amplitude was not imposed since the analysis of the observed amplitudes have shown that sometimes they tend to be larger than the real amplitude values due to observational scatter.   Moreover, RRab variables possessing periods in excess of one day are known to exist, yet such objects display smaller amplitudes in general and are therefore challenging to identify. The final suite of phased light curves \citep[following][]{dekany} were subsequently visually inspected (e.g., to ensure that they were asymmetric). \\

The final sample presented here includes 404 objects in total, spread over in 57\,sq.deg. Nearly 700 RRab stars were originally identified, and low quality candidates were subsequently culled based on their light curves.   Representative light curves of the sample are shown in Figure \ref{f:fig1}. Table \ref{t:tab1} (available in electronic form) provides for all 404 RRab variables IDs, coordinates, observed mean IR magnitudes and colors, periods, and amplitudes, as inferred from the $K_s$-band light curves.  The sample completeness is affected by numerous factors, such as: due to a magnitude-limited survey, faint and low amplitude variables are affected more severely by incompleteness; variables with symmetric light curve shapes may have been discarded to mitigate contamination from eclipsing binaries; Galactic location (varying field density and reddening); RR\,Lyrae located near the tile edges may possess larger uncertainties as the images are NIR mosaics; the total number of points per light curve varies between 50 and 130 epochs, therefore affecting the quality of the light curves; and the random temporal sampling of the light curves may introduce period aliases.  Note that there are typically 50 independent observations since the data are often acquired in pairs a few minutes apart. \\

\floattable
\begin{deluxetable}{cccccccccc}
\tablecaption{RR\,Lyrae sample photometric observations \label{t:tab1}}
\tablecolumns{7}
\tablenum{1}
\tablewidth{0pt}
\tablehead{
\colhead{VVV-RRL-} & \colhead{RA [hms]} & \colhead{DEC [dms]} & \colhead{L [deg]} & \colhead{B [deg]} & \colhead{$K_s$} & \colhead{$J-K_s$} & \colhead{$H-K_s$} & \colhead{P [days]} & \colhead{Ampl} \\
}
\startdata
 002 & 11:51:57.41 & -64:01:00.0 & 296.473 & -1.893 & 14.928 & 0.843 & 0.258 & 0.4567776 & 0.309 \\
 003 & 12:04:09.48 & -63:31:15.2 & 297.691 & -1.131 & 12.965 & 0.509 & 0.239 & 0.4509296 & 0.349 \\
 004 & 12:05:21.20 & -63:37:34.9 & 297.841 & -1.211 & 15.633 & 0.880 & 0.266 & 0.4128700 & 0.444 \\
 005 & 12:05:25.90 & -64:36:30.0 & 298.024 & -2.176 & 14.301 & 0.989 & 0.328 & 0.6404369 & 0.341 \\
 006 & 12:10:50.56 & -64:39:13.5 & 298.603 & -2.124 & 15.556 & 0.764 & 0.252 & 0.5541016 & 0.213 \\
 007 & 12:13:11.06 & -64:14:55.6 & 298.791 & -1.685 & 13.675 & 0.968 & 0.302 & 0.6429702 & 0.286 \\
 008 & 12:16:10.06 & -64:03:33.4 & 299.086 & -1.451 & 14.925 & 0.862 & 0.291 & 0.3751550 & 0.315 \\
 009 & 12:17:10.96 & -64:04:13.5 & 299.197 & -1.447 & 14.710 & 0.867 & 0.303 & 0.6037623 & 0.338 \\
 011 & 12:17:27.52 & -64:46:56.1 & 299.321 & -2.149 & 14.406 & 0.647 & 0.203 & 0.6163412 & 0.298 \\
 012 & 12:18:42.26 & -64:05:53.7 & 299.366 & -1.453 & 13.787 & 0.853 & 0.305 & 0.4762576 & 0.237 \\
\enddata
\tablecomments{Table \ref{t:tab1} is published in its entirety in the machine readable format.  A portion is
shown here for guidance regarding its form and content.}
\end{deluxetable}

\begin{figure}[h]
\centering
\includegraphics[width=0.6\hsize]{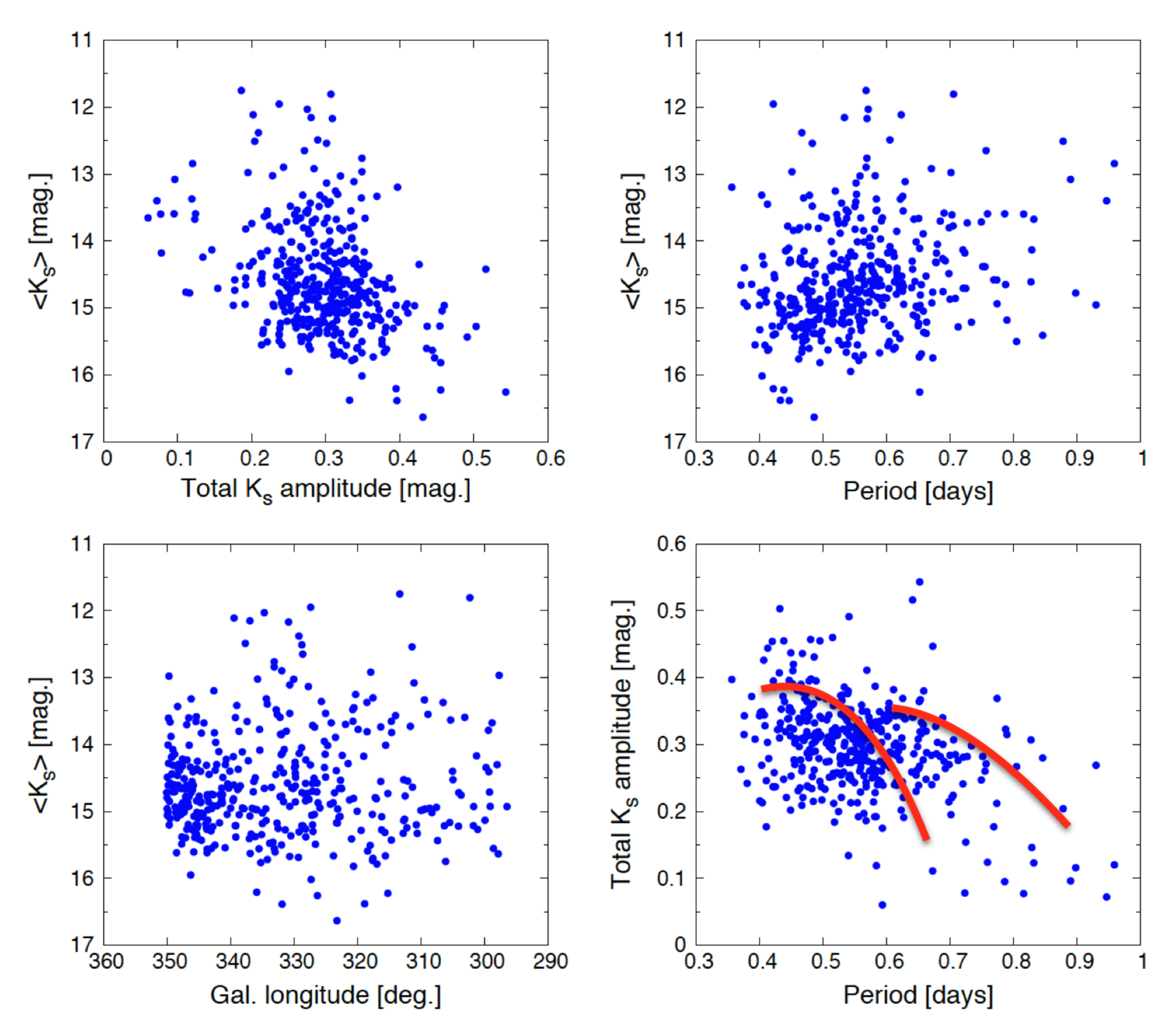}
\caption{Distribution of magnitudes, periods and amplitudes of the candidate RR\,Lyrae. Upper left: magnitudes vs amplitudes. Upper right: Magnitudes vs periods. Lower left: Magnitudes vs Galactic longitude. Lower right: Amplitudes vs periods (a.k.a. Bailey diagram). The sample RR\,Lyrae occupy the same regions in these diagrams as the bulge RR\,Lyrae discovered by Gran et al. (2016).  The main near-IR ridge lines for Oosterhoff I and II populations are plotted in red (right and left line, respectively) from Gran et al. (2016).  \label{f:fig2} }
\end{figure}

The distribution of observed parameters for the RRab sample are shown in Figure \ref{f:fig2} (i.e., locations, magnitudes, periods, and amplitudes).  Specifically, the figure conveys the sample's $K_s$-band magnitude as a function of the amplitude, Galactic longitude, and period, and a Bailey diagram (i.e., amplitude compared to the period). Our new sample of RR\,Lyrae in the disk occupies the same regions in these diagrams as the bulge RR\,Lyrae discovered by \citet{gran16}. The increased scatter observed arises principally from extreme reddening across the disk fields ($0.4<E(J-K_s)<3.5$), while the fields of the outer bulge exhibit $0.0 < E(J-K_s)<0.5$ \citep{gran16}.

\begin{figure}[h]
\centering
\includegraphics[width=0.7\hsize]{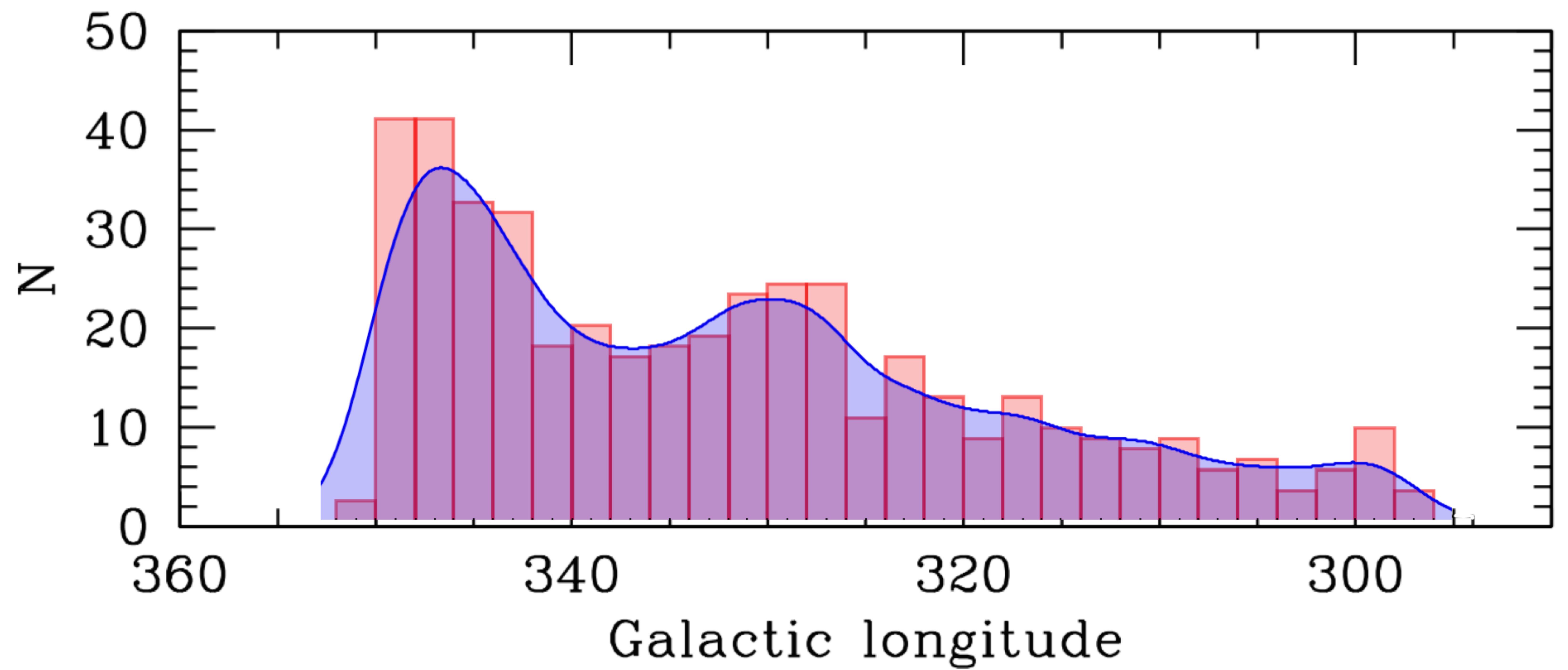}
\caption{Distribution in Galactic longitude of the newly discovered RR\,Lyrae type ab stars.  \label{f:fig3} }
\end{figure}

The discrete and smoothed distributions of the new sample of RRab stars as a function of Galactic longitude are shown in Figure \ref{f:fig3}. The sample spans from RA$=205\degr$ to $360 \degr$. The distribution expectedly increases towards the Galactic Center, given the higher stellar density in the bulge and Galactic Center \citep{valenti16}.   In addition, the observed distribution across the sky is inhomogeneous, displaying maxima and minima. \\

\begin{figure}[h]
\centering
\includegraphics[width=1.0\hsize]{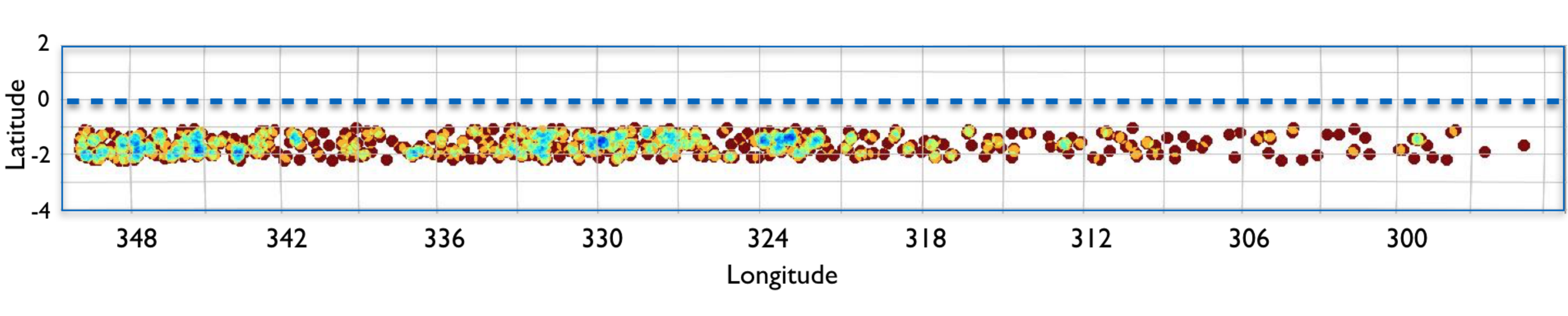}
\caption{Color surface density map showing the position of the candidate RR\,Lyrae in Galactic coordinates across 38 tiles at $b=-1.2$, covering 57 sq.deg. in total.  The different colors mark the density of overlapping points.} \label{f:fig4} 
\end{figure}

\begin{figure}[h]
\centering
\includegraphics[width=0.7\hsize]{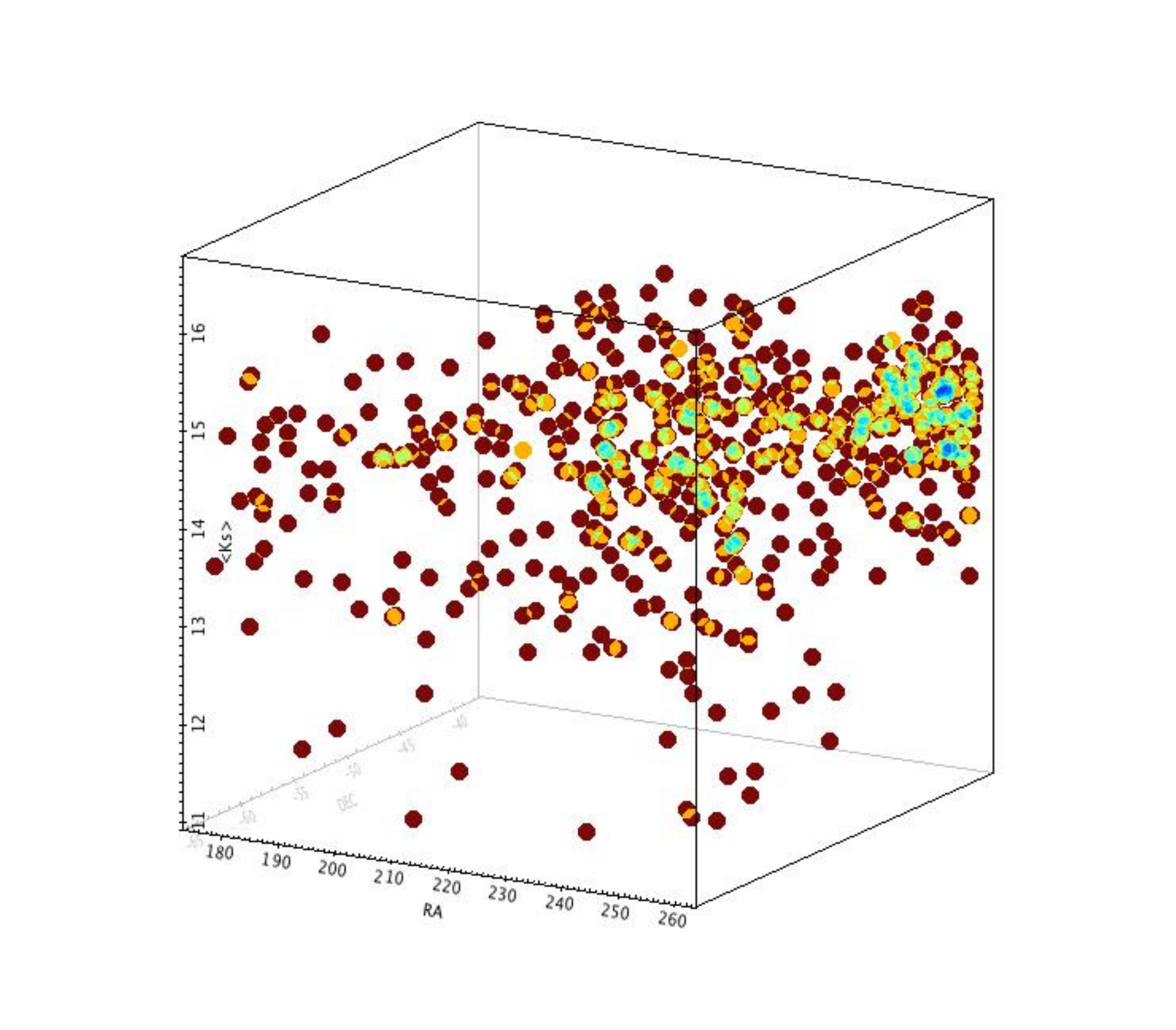}
\caption{3-D density map showing the position of the candidate RR\,Lyrae in equatorial coordinates across 38 tiles covering 57 sq.deg. in total. \label{f:fig5}}
\end{figure}

Figure \ref{f:fig4} is a surface density map detailing the positions of the RR\,Lyrae candidates in Galactic coordinates across 57\,sq.deg.\ within 38 tiles. The stars sample a narrow range of Galactic latitude below the plane, from $b=-2.24\degr$ to $b=-1.05 \degr$, while Figure \ref{f:fig5} conveys the 3-D density map of the candidates in equatorial coordinates relative to the $K_s$-band magnitudes.  The figures highlight the presence of groups of stars that may be associated, and voids are likewise apparent where few stars are located.  The spatial distribution is discussed further after the distance estimates are determined (Section~\ref{sec:distance}).\\

\section{The Color-Magnitude Diagrams} \label{sec:CMDs}
\begin{figure}[h]
\centering
\includegraphics[width=0.5\hsize]{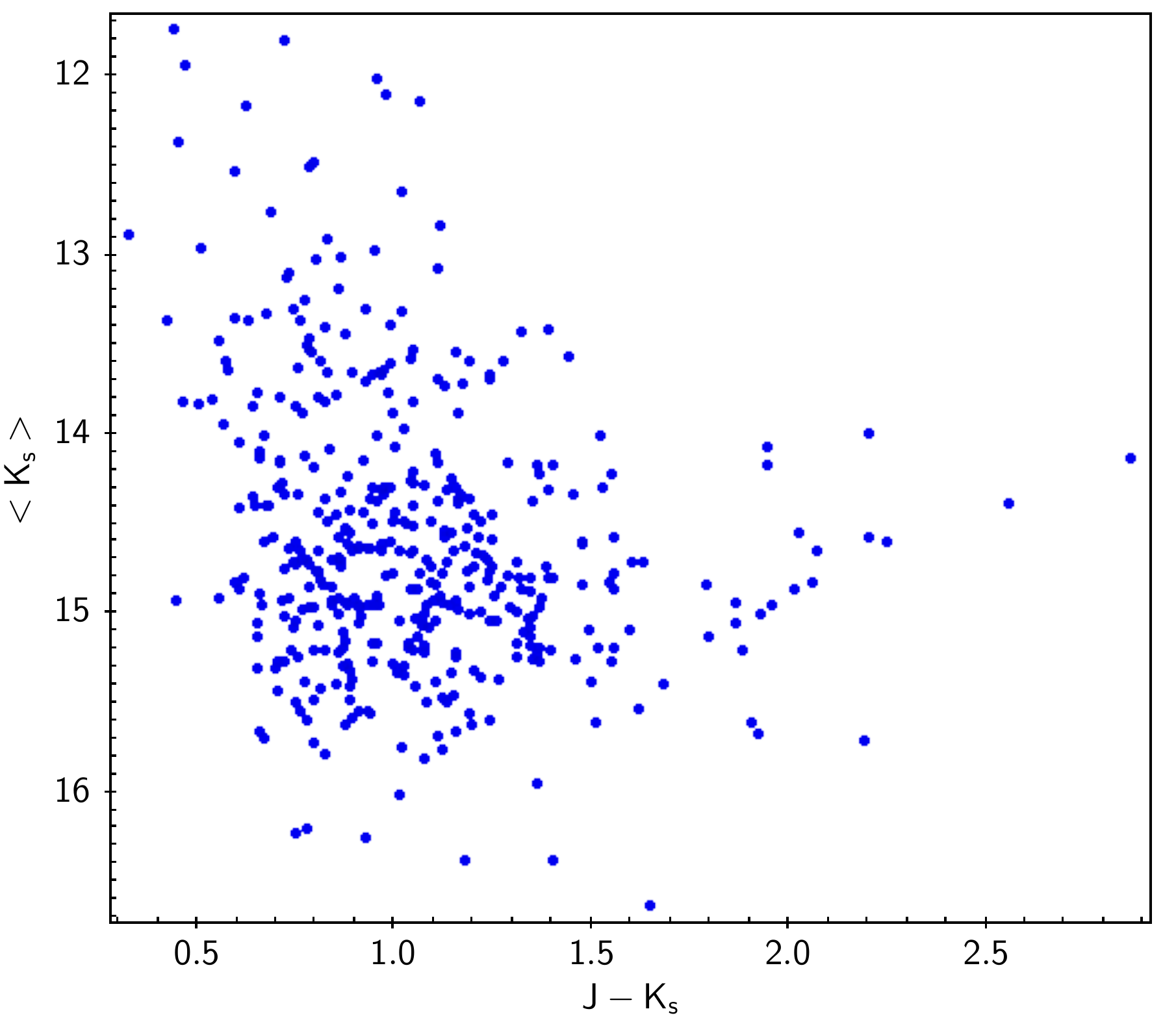}
\caption{Near-infrared color-magnitude diagram for the RR Lyrae candidates. These objects are much brighter than the limiting magnitudes in the tiles d001 to d038, that range from 17.5 to 18.5 mag in the Ks-band \citep{saito}. \label{f:fig6} }
\end{figure}

The NIR color-magnitude diagram (CMD) was constructed for the RR\,Lyrae candidates (Figure \ref{f:fig6}). The objects are significantly brighter than the limiting magnitudes associated with tiles d001 to d038, which range from 17.5 to 18.5 mag in the $K_s$-band \citep{saito}. The  RR\,Lyrae are typically fainter than bulge red clump giants, and are bluer than field clump giants.  Mean $K_s$-band magnitudes were utilized along with single-epoch $J$-band magnitudes to produce $(J-K_s)$, and consequently extra scatter is present due to random phase of J-band observations and periodic color variations. However, the amplitude changes are comparatively small in $(J-K_s)$, and are estimated to be $\Delta(J-K_s) < 0.1$ mag \citep{pietrzynski, borissova}, and the VVV photometric zero points are accurate to $0.02$ mag. Thus the principal scatter in color shown in the CMD arises from the extinction spread along the sight-lines to these low latitude RRab stars.  \\


A comparison of Figure \ref{f:fig6} with the CMDs of RR\,Lyrae in other regions of the Galaxy, such as the bulge, strengthens the case that the targets are RR\,Lyrae located at varying distances along the sight-line through the Galactic disk, with potentially large differences in extinction (see Section~\ref{sec:populations}).  \\

\begin{figure}[h]
\centering
\includegraphics[width=0.8\hsize]{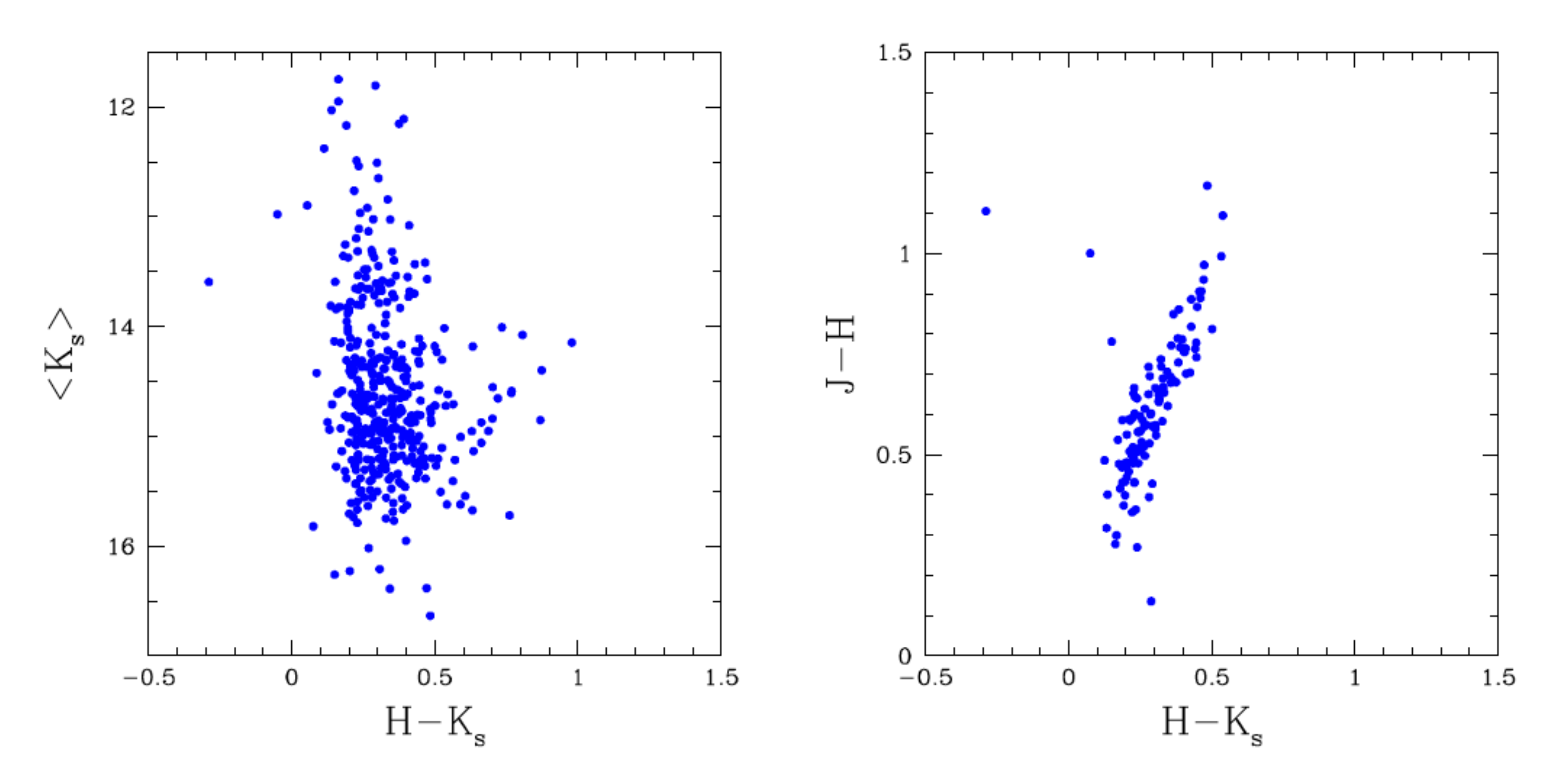}
\caption{Left panel: $K_s$ vs $(H-K_s)$ near-infrared color-magnitude diagram for the candidate RR Lyrae type ab stars. Right panel: $(J-H)$ vs $(H-K_s)$ near-infrared color-color diagram for the candidate RRab. The slope of the distribution in this color-color diagram agrees with the reddening vector from Nishiyama (2009). \label{f:fig7} }
\end{figure}

NIR color-magnitude ($K_s$ vs $(H-K_s)$) and color-color ($(J-H)$ vs $(H-K_s)$) diagrams were likewise constructed for the RRab candidate  sample (Figure \ref{f:fig7}).  The diagrams illustrate the variation in extinction as demonstrated by the extent of the data along the reddening vector. \\

\section{Reddenings and Extinctions} \label{sec:reddening}

Reddening and extinction values throughout the Galactic plane fields are extreme, and the uncertainties are exacerbated by possible variations in the extinction law \citep[for recent discussions see][]{nishiyama, gonzalez, chen13, nataf, majaess}. An important advantage of the $VVV$ photometry is that the impact of extinction is significantly reduced in the NIR with respect to the optical (by factors of $\sim$3 in reddening and $\sim$10 in extinction).\\

RR\,Lyrae stars are excellent reddening indicators because their intrinsic colors are known. They are located in the instability strip, which is defined by a narrow range of temperature and consequently intrinsic NIR colors within $\Delta (J-K_s)=0.10$ mag \citep[see][]{navarrete}.   Therefore, in principle, their observed colors can be used to estimate individual reddenings along the line of sight.  That property has been exploited to characterize the reddening throughout the bulge \citep[e.g.][]{alcock, popowski} based on optical data.  A mean intrinsic (i.e.\ unreddened) color of $(J-K_s) = 0.21 \pm 0.05$  is adopted for all RR\,Lyrae stars in our sample (see \citet{gran16})\footnote{There is no concensus about this mean value \citep[e.g.][]{navarrete, gran15, klein+bloom14, muraveva}, but the scatter in this color is so small that in no case changes significantly the conclusions of this work.}. Furthermore, as mentioned earlier, the $(J-K_s)$ color range throughout the phase cycle is small, $\Delta(J-K_s)<0.1$ mag. Exceptions to the adopted mean parameters may be present for RR\,Lyrae that are blended with field stars. The impact of blending would be bluer observed colors if the blend is with a foreground disk main-sequence star, or redder observed colors if the blend is with a background red giant.  The effect of blending is mitigated since the VVV images exhibit high resolution ($\sim$ 0.9 arcsec), and consequently blends should be rare because the stellar densities at the relevant magnitudes are not very high even in the low latitude fields observed here. Moreover,  the impact of the Blazko effect is reduced in the NIR, and in particular the mean colors are unaffected\footnote{The Blazko effect  produces varying amplitudes for RRab stars \citep{blazko}, and in the optical the colors are often measured at minimum light where they are more stable \citep{kunder10}.}.\\

Figure \ref{f:fig7} illustrates the significant variation in reddening, as RR\,Lyrae possessing intrinsic colors $(H-K_s) = 0.05 \pm 0.02$ display observed colors spanning from $(H-K_s) \sim 0.0$ to $(H-K_s)\sim 1.0$, and beyond. Certain outliers may be blended stars.  The choice of the reddening law is important, and can impact the computed distances. That uncertainty is apparent when comparing the extinction based on reddening laws of \citet{nishiyama} with \citet{cardelli}. The extinction is $A_{K_s} =  0.528\,E(J-K_s)$ for Nishiyama, and $A_{K_s} = 0.72\,E(J-K_s)$ for Cardelli law, respectively.
There is a systematic effect on the distances depending on the PL relation and the reddening law adopted. It is straightforward to consider the effect of the reddening curve on the distances. The steeper the adopted extinction curve, the smaller the absolute-to-selective extinction ratio, thus smaller the absolute extinction for the same measured color excess. Therefore, if one computes distances adopting a steeper extinction curve, one will get larger distances for the objects for the same color excess values (E(J-Ks) in our case). If the assumption about the validity of the Nishiyama extinction curve is incorrect  for the present sample, and our objects follow a shallower, more standard extinction curve, then the distances would get systematically shorter. \\

In sum, the reddening was determined via the following relation: $E(J-K_s) = (J-K_s) - 0.21$, where 0.21 mag is the intrinsic $(J-K_s)$ color for RRab stars. The following expression was adopted to convert the reddening to the total extinction: $A_{K_s}=0.528\,E(J-K_s)$, because the reddening vector defined by the RRab distribution in the color-color diagram (Figure \ref{f:fig7}) matches the Nishiyama (2009) result, namely $E(J-H)/E(H-K_s) = 2.05$.  The sample exhibits a reddening baseline ranging from approximately $E(J-K_s)=0.4$ to $3.0$. The equivalent extinction values in the $K_s$-band span $0.2<A_{K_s}<1.6$ mag, and in the $V$-band that implies $1.8<A_V<14.6$ mag. The typical reddening value for the sample is $E(J-K_s)=1.0$ mag, which is equivalent to $A_{K_s}=0.5$ mag and $A_V=4.8$ mag.\\


Knowledge of the extinction values and the slope of the reddening law may be utilized to estimate the expected visual magnitudes for the RR Lyrae variables discovered. The objective is to evaluate the possibility of undertaking follow-up observations for the RR\,Lyrae sample in the optical, and specifically, if the objects will be surveyed by Gaia.  The Gaia mission will revolutionize the study of Galactic structure in general, and of RR\,Lyrae as distance indicators in particular \citep[e.g.][]{wilkinson, bailer, eyer, clementini}. \\ 

\begin{figure}[h]
\centering
\includegraphics[width=0.4\hsize]{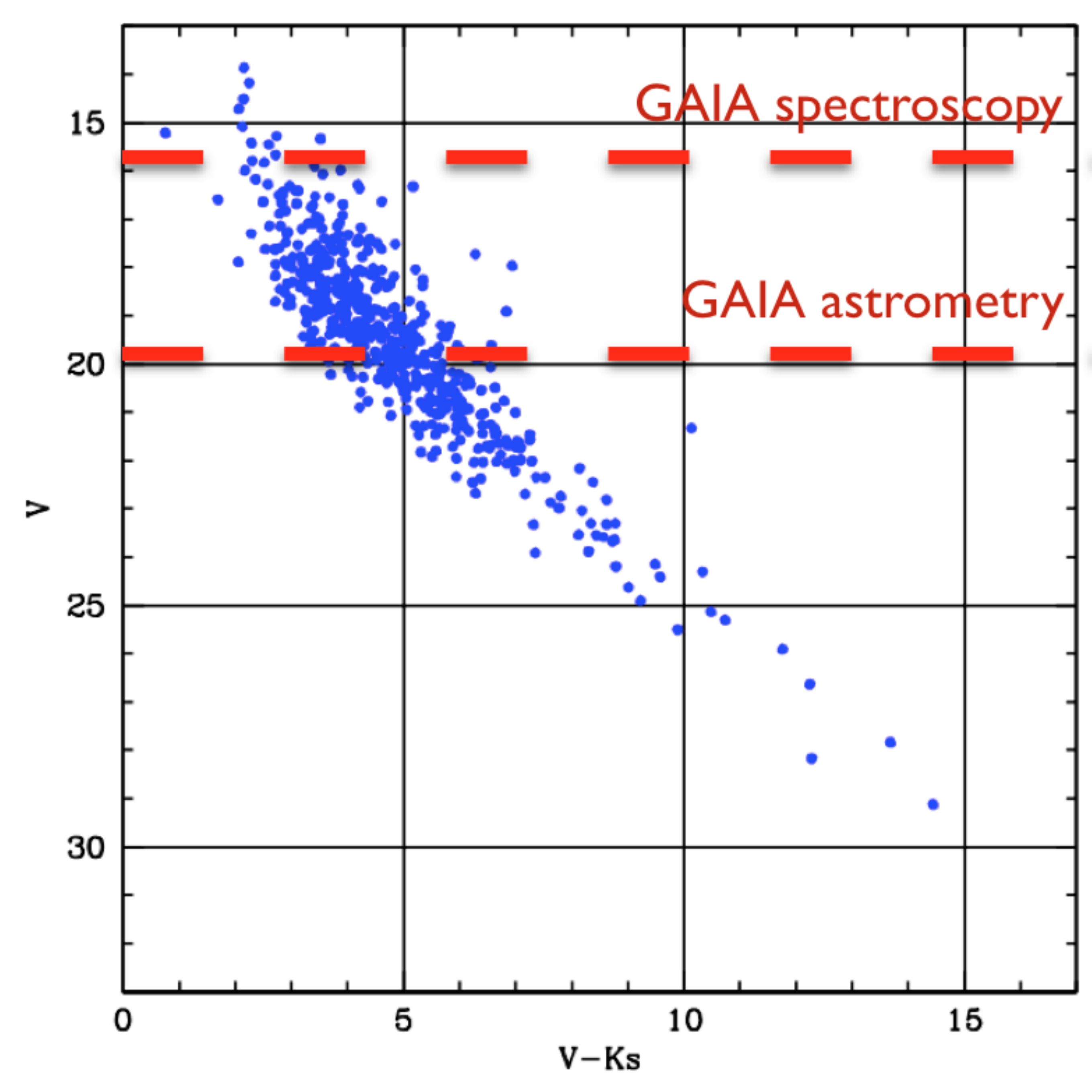}
\caption{Color-magnitude diagram for the sample RR\,Lyrae assuming visual magnitudes computed as explained in the text. The limits for Gaia astrometric and spectroscopic observations are shown for comparison. \label{f:fig8} }
\end{figure}

A mock optical-infrared $V$ vs $(V-K_s)$ CMD (Figure \ref{f:fig8}) was constructed as follows. Mean colors for unreddened Galactic RR\,Lyrae ($E(B-V) \sim 0.015$) were adopted from \citet{muraveva}. The intrinsic $(V-K_s)$ color for RRab is $M_V-M_{K_s} = 1.006$, and therefore the $K_s$-band magnitude corrected for extinction is ${K_s}_0 = K_s - A_{K_s} = K_s - 0.528\,E(J-K_s)$ (for the \citet{nishiyama} reddening law). The visual magnitudes were subsequently estimated via: $V_0= K_s - 0.528\,((J-K_s) - 0.21) + 1.0$, where $(J-K_s)_0 = 0.21$ is assumed to be the color of an unreddened RR\,Lyrae star.
The visual magnitude follows from the intrinsic color, extinction, and distance. We indicate the estimated limits for Gaia astrometric and spectroscopic observations \citep{wilkinson, clementini} on the CMD in Figure \ref{f:fig8}. Numerous RR\,Lyrae stars ($\sim$ half of our sample) have $V < 20$ mag, and are therefore accessible to Gaia astrometric observations. Moreover, about 10 have visual magnitude within reach of the spectroscopic observations with Gaia.  Even though Gaia will provide useful proper-motions, no useful ($5 \sigma$)  parallaxes are expected for $V>15$ mag sources that are more than 8 kpc away. \\

\section{Distances} \label{sec:distance}

In this section we derive individual distances for our RR\,Lyrae sample.  There are several factors to consider when estimating such distances, beyond the already mentioned issues of high extinction and an uncertain extinction law. Chief among them are the period-luminosity (PL) relations and absolute magnitudes, and their potential dependence on metallicity.\\

\citet{muraveva} presented recently new near-IR PL and period-luminosity-metallicity (PL$_{\mathrm{K_s}}$Z) relations for RR Lyrae stars, and compared them with the theoretical and empirical relations from the literature \citep{bono03,catelan04,dallora04,delprincipe06,sollima06,sollima08,borissova}. Based on this we assume $0.1$ mag as an upper limit to the uncertainty for the PL relation when computing the distances.  The \citet{muraveva} PL$_{\mathrm{K_s}}$ relation is tied to five RR\,Lyrae with HST-based parallaxes \citep{benedict}. Distances for our new RRab sample were derived using the PL relation given by Eq.~14 of \citet{muraveva}, that is:  $M_{K_s} = -2.53\,log(P) - 0.95$.  A period-luminosity-metallicity relation is likewise presented in that work, but the present RR\,Lyrae sample lacks accurate individual metallicities. While later we use the periods to determine the metallicity distribution, individual metallicities determined in this fashion do not provide improved distance estimates. \\

\begin{figure}[h]
\centering
\includegraphics[width=0.5\hsize]{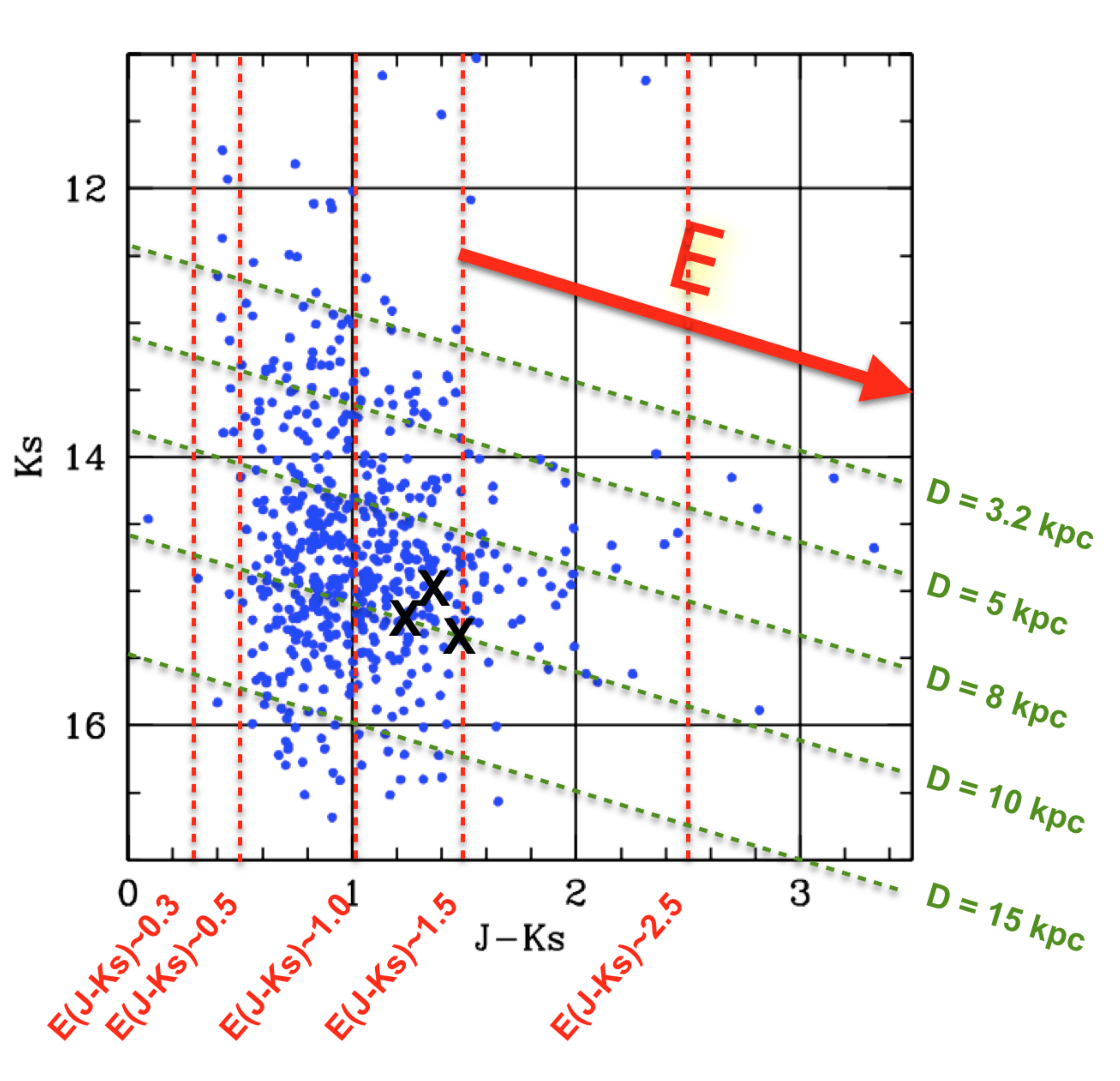}
\caption{Near-infrared color-magnitude diagram for the candidate RR\,Lyrae showing lines of constant reddenings and distances. We have assumed the slope of the reddening vector from \citet{nishiyama} as shown. Fiducial reddening and distance lines for the RR Lyrae are marked with dashed red and green lines, as labelled. The X mark the position of two RRab which belong to the globular cluster 2MASS-GC03 (FSR\,1735) \citep[see][]{carballo}.  \label{f:fig9}
}
\end{figure}

A NIR CMD for the RRab sample overlaid with lines of constant reddening and distance is presented in Figure \ref{f:fig9}.  The diagram illustrates the range of distances and extinctions, indicating that our sample becomes incomplete for $D > 20$ kpc, and reddenings and extinctions of $E(J-K_s) > 2.5$ and $A_{K_s} > 1.2 $ mag, respectively.  The lines of constant distance were inferred from $P\sim 0.7$ day RRab stars, and targets with different periods will scatter around those representative lines. The CMD excludes 14 stars that lack $J$-band observations,
and the reddening vector from \citet{nishiyama} is shown. \\

There are numerous groups (pairs, triplets) of RRab variables within the discovered sample that lie in close spatial proximity. It is worth following up on these groups as, for example, they may potentially be associated with tidal streams or globular clusters.  Three  RRab stars appear to belong to the globular cluster 2MASS-GC03 (FSR\,1735), located in the VVV tile d031 at RA=16:52:11.1, DEC=--47:03:23, or $l=339.18979$,  $b=-1.85318$. The globular cluster provides a valuable anchor to check the computed distances. The positions of the three RRab members of this cluster \citep[V1, V2 and V3 in the notation of ][]{carballo} are highlighted in Figure \ref{f:fig9}. \citet{carballo} measured a mean cluster metallicity and radial velocity of $[Fe/H] = -0 .9 \pm 0 .2$ and $RV= -78 \pm 12$ km/s, respectively.  Our estimated mean distance to the RR\,Lyrae is $D = 10.8 \pm 0.4$ kpc.  That distance is in agreement with their position in the cluster CMD \citep{carballo}. The mean reddening and extinction inferred from the RR\,Lyrae stars are $E(J-K_s) = 1.3 \pm 0.1$,  $A_{K_s} = 0.69$ mag, and $A_V= 6.3 $ mag, accordingly. This demonstrates that our distances are reasonable, and justifies the selection of the reddening law, absolute RR\,Lyrae $M_{K_s}$, and PL relation.  In addition, another group of disk RR Lyrae points to the discovery of a new globular cluster in the Milky Way, which we name VVV-GC-05 (FSR1716), and study in detail in a follow-up paper (Minniti et al. 2017). \\


Our RR\,Lyrae sample is located in the direction of the thin disk ($-2.24<b<-1.05$ deg), and complements previous surveys associated with the thick disk, halo, and the well-studied local sample of \citet{layden94}. The new RR\,Lyrae are concentrated toward the Galactic plane, with $0 < z < -850$ pc, and exhibit a mean height below the Galactic plane of $-240$ pc. \\

\begin{figure}[h]
\centering
\includegraphics[width=0.65\hsize]{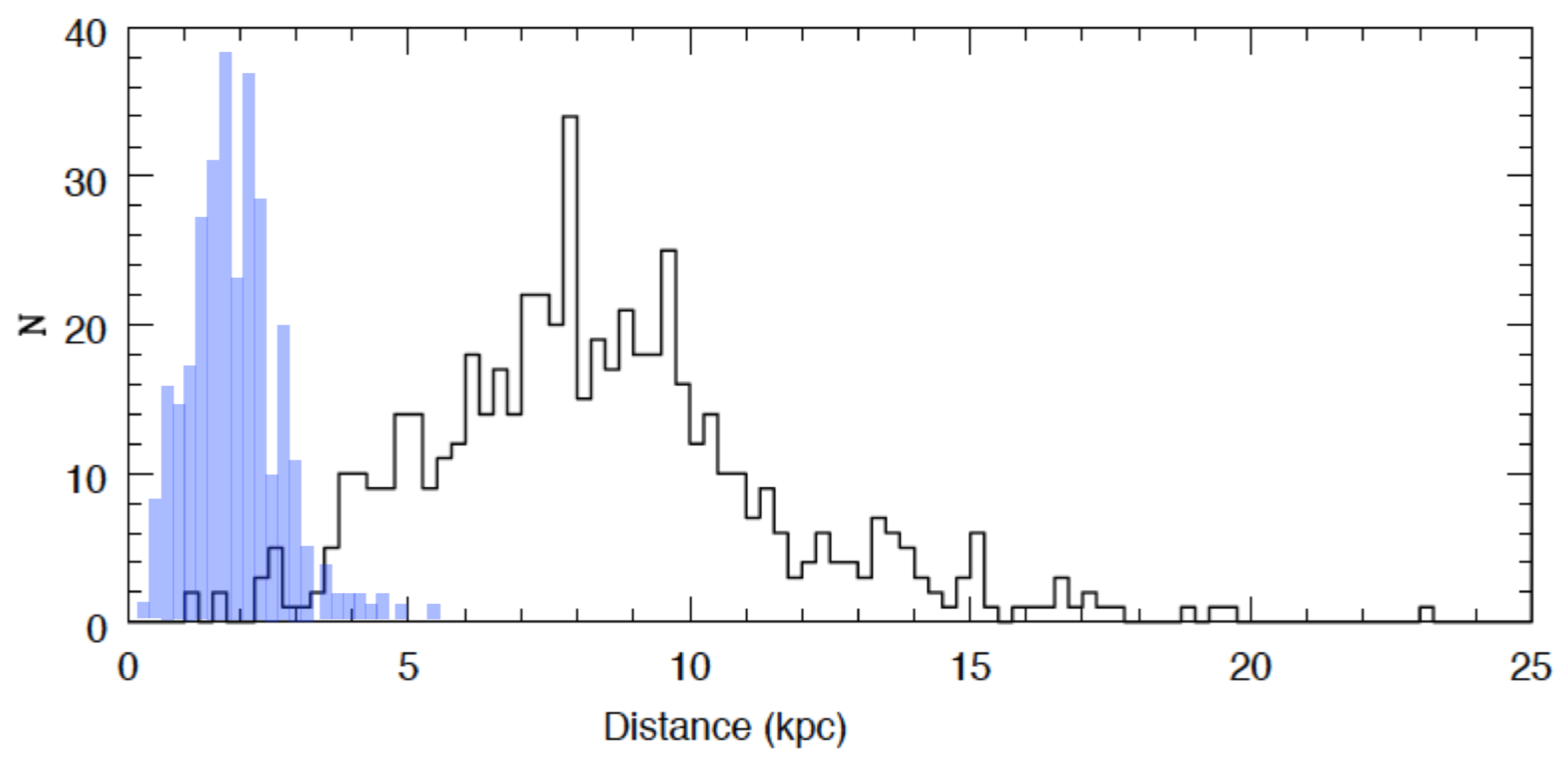}
\caption{Distance distribution for the RR Lyrae along the line of sight. This is shown with a small bin size (0.25 kpc), in order to illustrate possible groupings along the line of sight. This sample is very complementary to the local sample of \citet{layden94}, that contains RR\,Lyrae with D\,$<5.5$ kpc (shown in light blue). \label{f:fig10} }
\end{figure}

The distance distribution for the RRab sample along the line of sight is shown in Figure \ref{f:fig10}.  The density distribution peaks at a distance of $D \sim 8$ kpc, which is where the RR\,Lyrae are closest to the Galactic center and it is the tangent point with the Galactic bulge. Importantly, the RR\,Lyrae distribution does not peak at the end of the bulge bar, which is located at a distance of $D = 13-14$ kpc. The result supports claims that the RR\,Lyrae distribution is not barred \citep{dekany} \citet{gran16}.  An excess number of RR\,Lyrae lies between $6$ and $10$ kpc, which is the contribution of the bulge demographic. \\

The distance distribution in Figure \ref{f:fig10} appears inhomogeneous, and is potentially revealing true substructure. The distribution is portrayed using a small bin size (0.25 kpc) to illustrate possible groupings along the sight-line. For example, excess RR\,Lyrae are observed toward tile d026 (centered at  $l= 332$, $b=-1.7$), and the density enhancement warrants further study.  If the observed substructure(s) are not spurious artefacts of small statistics, two separate conclusions may be considered: (1) the distribution may be intrinsically inhomogeneous (e.g. due to halo streams, previously unseen globular clusters or dwarf galaxies, massive disk spiral arms, bulge orbital traps, etc.), (2) the observed distribution is biased by high extinction, as illustrated when comparing the observed spatial distribution with extinction maps (see Figure 4).\\

\begin{figure}[h]
\centering
\includegraphics[width=0.7\hsize]{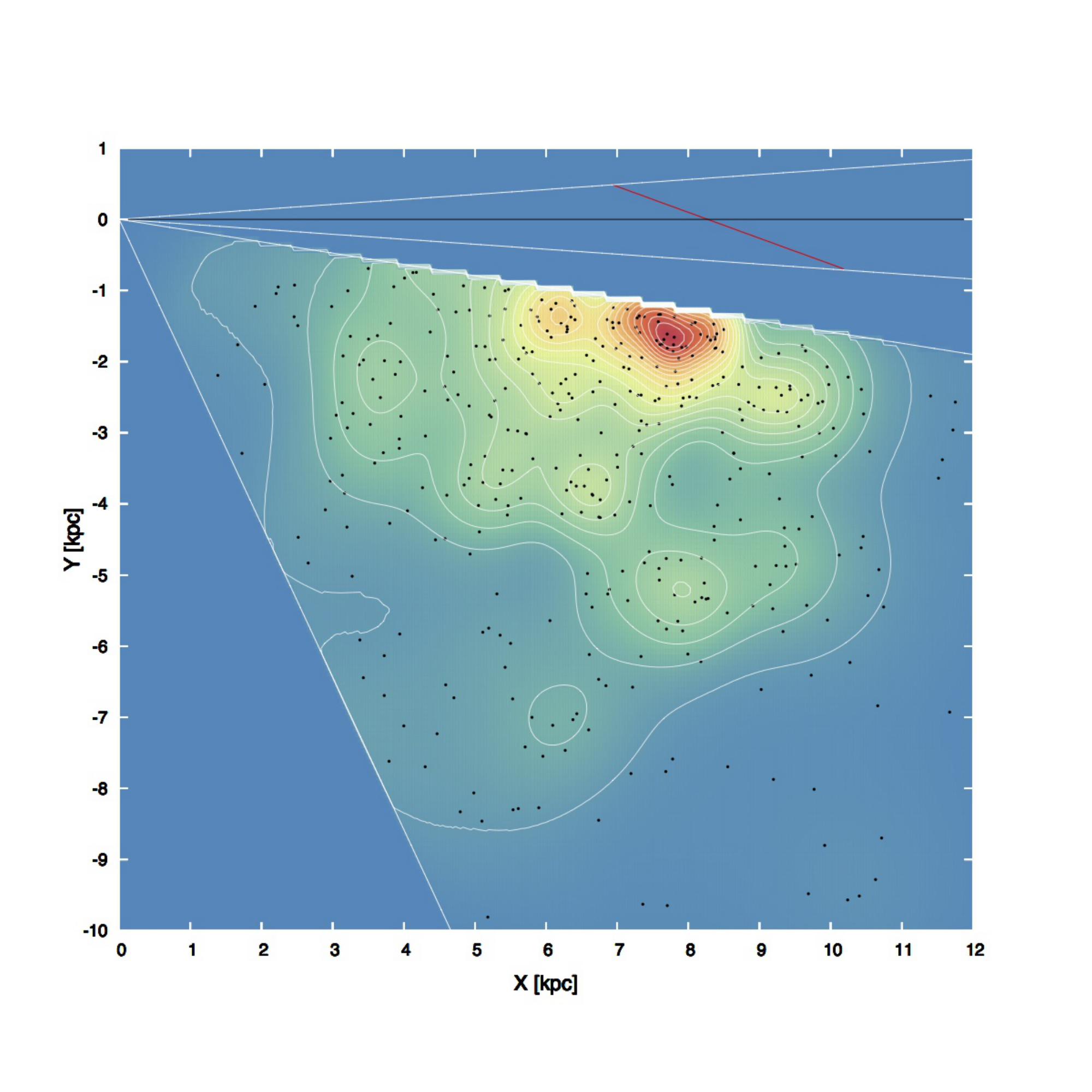}
\caption{Face-on density map, where the maximum density occurs at the bulge tangent point, as expected if the RRab distribution is not barred but rather spherically symmetric \citep{dekany, gran16}. The spatial substructure of the sample is also apparent. Black dots represent the detected RR\,Lyr positions. The longitudinal boundaries of the current study and those of \citet{pietrukowicz15} are shown with white lines.  also the position of the long axis of the ellipse fitted by these authors is shown in red line (with  inclination of 20 deg and Galactic center distance of 8.27 kpc, both from their analysis). The Galactic center direction is also shown with a black line.} \label{f:fig11} 
\end{figure}

Figure \ref{f:fig11} conveys a map of the Milky Way together with the distribution of the RRab stars. 
The figure demonstrates that the peak density is located at the bulge tangent point, as expected if the RRab distribution is not barred, but rather axisymmetric \citet{dekany} \citet{gran16}.  Furthermore, the bulge RR Lyrae population appears to extend from $\ell \sim 344 \degr$, beyond which the excess defined by the bulge RR\,Lyrae disappears (as seen in Figure \ref{f:fig4}). Expectedly, Figure \ref{f:fig11} likewise shows that the substructures do not trace the spiral arms, and are instead described by a smoother distribution across the Galaxy.  Interestingly, certain distinct substructures warrant further investigation. For example, there is a dearth of RR\,Lyrae in a region of the Galactic plane beyond 10 kpc for longitudes $332 < \ell < 340 \degr$. The absence of remote RRab stars in that region may stem from the presence of a distant thick interstellar cloud producing extreme extinction.  In sum, Figure \ref{f:fig11} conveys density enhancements beyond the bulge, which may belong to specific Galactic components (i.e.\ the disk or the halo). \\

\section{Metallicities} \label{sec:metallicity}

\begin{figure}[h]
\centering
\includegraphics[width=0.5\hsize]{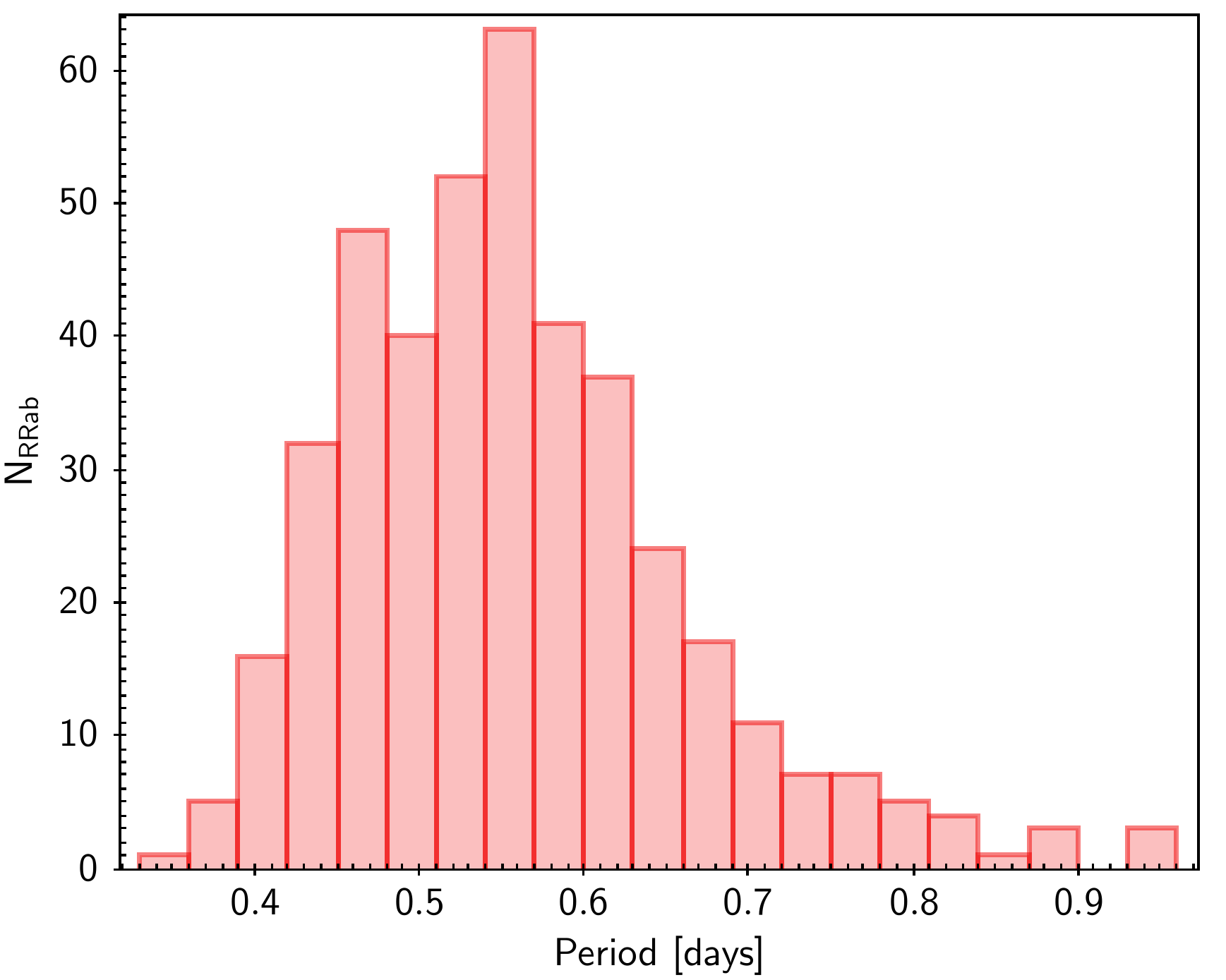}
\caption{Period distribution for the RR\,Lyrae sample. The mean of the distribution corresponds to an Oosterhoff type I population (see text).  \label{f:fig12} }
\end{figure}

A plethora of RR\,Lyrae exists in the halo fields of the Milky Way, and the stars are likewise often present in sizable numbers in Galactic globular clusters. However, not all globular clusters contain RR\,Lyrae, and the variables are more common in metal-poor clusters.  The observed period distribution for the newly discovered sample of RRab stars exhibits a mean period of $P = 0.5586$ days (Figure \ref{f:fig12}), which indicates that the majority of the RRab resemble an Oosterhoff type I population. RR\,Lyrae are often divided into two main populations designated Oosterhoff type I or type II \citep{oosterhoff, kinman, catelan}. The Oosterhoff populations have different period distributions in globular clusters relative to the halo fields, with the Oosterhoff type I populations displaying shorter mean periods ($P < 0.6$ days) and higher metallicities ($[Fe/H] > -1.5$ dex) than their type II counterparts \citep{catelan}.\\

\begin{figure}[h]
\centering
\includegraphics[width=0.6\hsize]{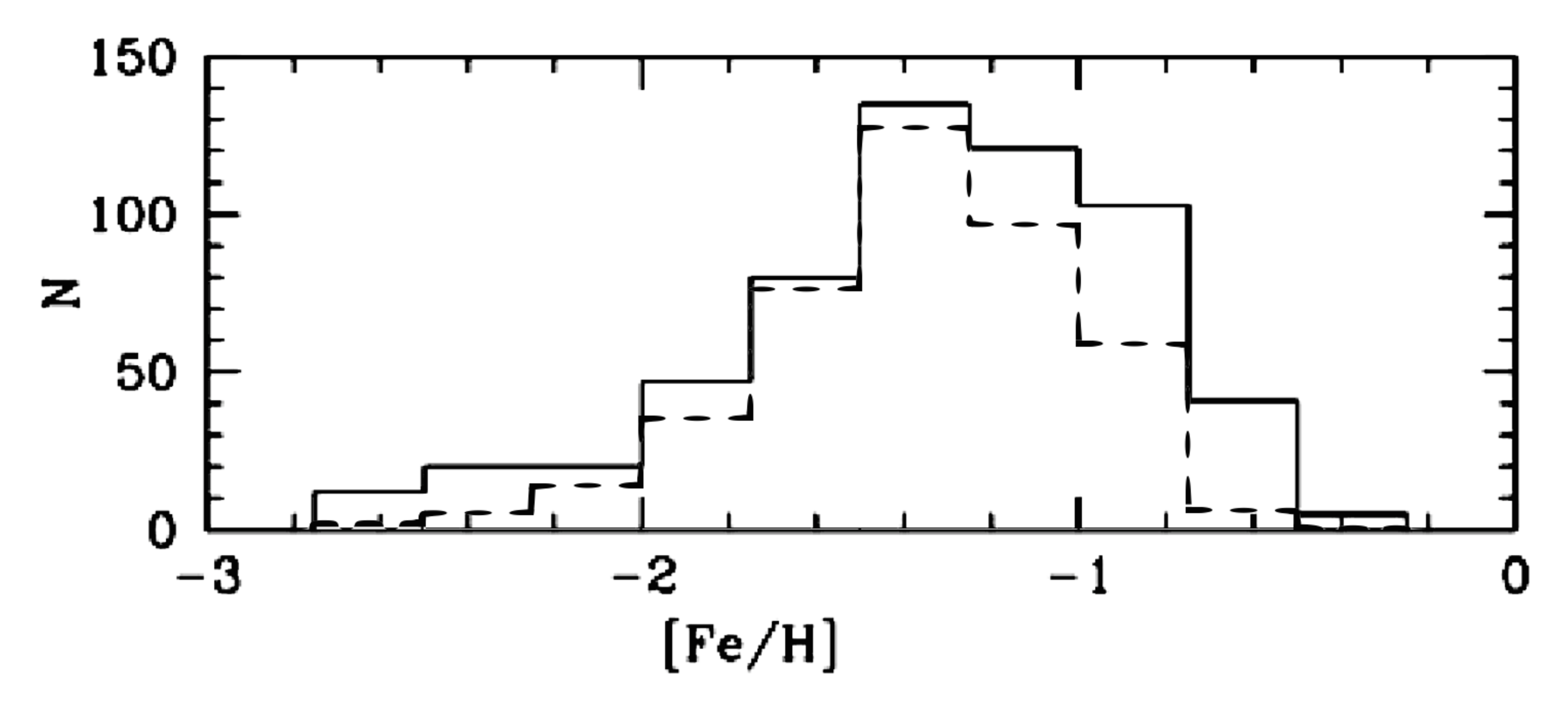}
\caption{Metallicity distribution for the sample RR Lyrae (black solid line) compared with the sample of Gran et al. (2016) arbitrarily normalized (dashed line). These metallicities were estimated using the periods \citep{yang,feast}, and should be only considered as rough estimates until spectroscopic determinations become available.  \label{f:fig13} }
\end{figure}

Individual metallicities for RRab stars may be estimated from their periods, with reference to the RR\,Lyrae calibrators of \citet{layden94}. \citet{feast}, and \citet{yang} cite the following relations: $[Fe/H] = -5.62\,\log P_{ab} - 2.81$, $\sigma = 0.42$, and $[Fe/H] = -7.82\,\log P_{ab} - 3.43$, $\sigma = 0.45$, respectively. The derived metallicity distribution for the RR\,Lyrae sample using the \citet{feast} relation yields a mean metallicity of $[Fe/H] = -1.3 \pm 0.2$ dex (Figure \ref{f:fig13}). The individual metallicities were estimated using the periods, and are merely estimates (see discussions by \citet{yang}, and \citet{feast}) in the absence of spectroscopic determinations. The individual metallicities may have a scatter of several dex, yet the shape of the metallicity distribution confirms that the RRab sample constitutes metal-poor objects which belong predominantly to an Oosterhoff type I population.  \\

Figure 13 shows the metallicity distribution of the present RR Lyrae sample compared with the RR Lyrae  population in the outer bulge from Gran et al. (2016). While the peak locations are similar, the present metallicity distribution is broader than that of the outer bulge sample. This figure shows also that there is a lack of very metal-rich RRab (having up to Solar metallicity), that are seen in the local populations (\citet{layden94}), suggesting that our sample is dominated by halo RRabs, containing few disk RRab. We predict that the present relatively metal-poor sample should exhibit high velocity dispersion. This needs to be explored spectroscopically, by measuring accurate radial velocities and metallicities. \\

\section{Comparison with other RR Lyrae populations} \label{sec:populations}

The properties for all targets are listed in Table \ref{t:tab2} (available in electronic form), which include the reddenings, distances, and metallicities.  The RR\,Lyrae sample discovered is now compared to several results, namely the inner and outer bulge RRab, globular cluster $\omega$\,Cen RRab members, the \citet{layden94} local sample, and the halo population.  \\

(1)	the inner bulge RRab:\\
\noindent \citet{dekany} studied the inner bulge RR\,Lyrae stars and established accurate distances.  They concluded that the RR\,Lyrae distribution is axisymmetric, and not barred. \citet{pietrukowicz15} utilized optical photometry from OGLE and contested that result, since they discovered the population adhered to a barred distribution. However, the spatial density distribution of our sample does not peak near the bar's end ($D \sim 13-14$ kpc), but the maximum is instead observed at the bulge tangent point near $D \sim 8$ kpc, as expected from an axisymmetric distribution. \\  

(2)	the outer bulge RRab:  \\
\noindent \citet{gran16} showed that the outer bulge RR\,Lyrae do not trace a bar component, whereas a well-defined bar signature is delineated by red clump giants. They argued that the RR\,Lyrae in the bulge-halo transition have an axisymmetric distribution, which supporting the results of \citet{dekany}.  \\

(3)	the globular cluster $\omega$\,Cen: \\
\noindent \citet{navarrete} recently completed a NIR analysis of the RR\,Lyrae population associated with $\omega$\,Cen, the most massive Galactic globular cluster, which is considered to be the remnant nucleus of a disrupted dwarf galaxy. $\omega$\,Cen is the prototypical example of an Oosterhoff type II population, while the new RR\,Lyrae sample appears to be a composite population with a large metallicity spread, but dominated by an Oosterhoff type I population, which features a more metal-rich mean ($[Fe/H] = -1.3$).  Only $\sim 30\%$ of the sample are of the Oosterhoff type II class like the globular cluster $\omega$\,Cen, assuming an arbitrary division of $[Fe/H] = -1.5$ for the separation between the Oosterhoff classes. \\

(4)	the local sample: \\
\noindent The local sample is likewise a composite Oosterhoff population, but dominated by the Oosterhoff class I RR\,Lyrae \citep{layden94}. These RR\,Lyrae are located within 5 kpc from the Sun, and appear to be representative of the population found across the plane, since there are no marked differences with remote RR\,Lyrae discovered in the present study. The new RR\,Lyrae exhibit Galactocentric distances spanning $R = 1.4-20$ kpc, which extends and complements the \citet{layden94} sample, which ranges $R = 6-13$ kpc. The present sample features RR Lyrae at smaller Galactocentric distances, and includes RR\,Lyrae belonging to the outer bulge population, but ultimately the properties are similar to those of local RR\,Lyrae.   \\
 
(5)	the halo population: \\
\noindent There is considerable research concerning the halo RR\,Lyrae population, because it is a useful tracer of the halo's substructure \citep[e.g.][]{vivas,keller,miceli,catelan,sesar,baker}. Certain RRab stars in our sample may belong to a halo field population traversing the Galactic plane during their orbits, and under that scenario they should possess high proper motions. Alternatively, if the RR\,Lyrae are endemic to the Galactic disk, an implication is that the disk is extremely old ($ > 10$\,Gyr), and must have been \textit{in situ} since very early in the Milky Way's formation. Follow-up by Gaia is important in order to establish the kinematics of the RR Lyrae sample, and potential membership within the halo population.
In addition, complementary NIR spectroscopy may provide more accurate individual metallicities, and also radial velocities, in order to infer mean orbital properties ($V_{rotation}$, and line of sight velocity dispersion, Frenk \& White 1980), that would help to clarify disk vs halo vs bulge membership. \\


\floattable
\begin{deluxetable}{ccccc}
\tablecaption{Measured stellar parameters\label{t:tab2}}
\tablecolumns{6}
\tablenum{2}
\tablewidth{0pt}
\tablehead{
\colhead{VVV-RRL-} & \colhead{$E(J-K_s)$} & \colhead{$A_{K_s}$} & \colhead{D [kpc]}  & \colhead{$[Fe/H]$} \\
}
\startdata
   002 &  0.693 & 0.366 &  8.516  & -0.9  \\
   003 &  0.359 & 0.190 &  3.716  & -0.9  \\
   004 &  0.730 & 0.730 &  11.095 & -0.6   \\
   005 &  0.839 & 0.443 & 7.306  & -1.7  \\
   006 &  0.614 & 0.324 &  12.783 & -1.4    \\
   007 &  0.818 & 0.432 & 5.515 & -1.7    \\
   008 &  0.712 & 0.376 &  7.662 & -0.4  \\
   009 &  0.717 & 0.378 &  8.819 & -1.6   \\
   011 &  0.497 & 0.262 &  8.173  & -1.6  \\
  012 &  0.703 & 0.371 & 5.130 & -1.0   \\
\enddata
\tablecomments{Table \ref{t:tab2} is published in its entirety in the machine readable format.  A portion is
shown here for guidance regarding its form and content.}
\end{deluxetable}

\section{Conclusions}\label{sec:conclusion}
We have discovered 404 RRab candidate variable stars across a 57\,sq.deg. region along the southern plane of the Milky Way.  NIR photometry from the $VVV$ survey was employed to establish accurate positions, near-IR magnitudes, colors, periods, and amplitudes as well as to estimate metallicities for the whole sample. \\

The RRab stars are located at distances spanning from $1 < D < 30$\,kpc, with the maximum density being at the distance of the Galactic bulge. This indicates that the RRab distribution does not trace the bar delineated by red clump giants, while it is compatible with an axisymmetric distribution \citep{dekany}. Most of the stars have properties indicative of an Oosterhoff type I population.  The spatial substructure exhibited in the present sample may be real and warrants further study. For example, three RR\,Lyrae appear to belong to the Galactic globular cluster 2MASS-GC03 (FSR1735), other five RRab trace a new globular cluster (VVV-GC05, FSR1716; Minniti et al. 2017), and additional groupings may trace other substructures. While a sizable fraction, close to the Galactic Center belongs to the bulge RR\,Lyrae population, those further away may belong to a high proper motion halo field population traversing the Galactic plane. These findings may support results from hydrodynamical cosmological simulations where stars formed in satellite galaxies are later on accreted to form part of the inner stellar halo and bulge (Tissera et al. 2017) in Milky Way mass-like galaxies. Alternatively, the RR\,Lyrae may belong to the Galactic disk, which would imply the disk is old ($>10$\,Gyr) and must have been in place since very early in the Galaxy's formation. Additional observations are necessary to select a scenario, and specifically, proper motions are particularly desirable. The RR\,Lyrae discovered here are also prime targets for near-IR spectroscopic observations, namely to measure chemical compositions and radial velocities. Lastly, our brightest cataloged RR\,Lyrae constitute a prime target list for the Gaia mission. \\

\acknowledgments
We gratefully acknowledge the use of data from the ESO Public Survey program ID 179.B-2002 taken with the VISTA telescope, and data products from the Cambridge Astronomical Survey Unit (CASU).  Support for the authors is provided by the BASAL Center for Astrophysics and Associated Technologies (CATA) through grant PFB-06, and the Ministry for the Economy, Development, and Tourism, Programa Iniciativa Cientifica Milenio through grant IC120009, awarded to the Millennium Institute of Astrophysics (MAS).  D.M., J.A.G. and M.Z. acknowledge support from FONDECYT Regular grants No. 1130196, 11150916 and 1150345, respectively. R.K.S. acknowledges support from CNPq/Brazil through project310636/2013-2. We are also grateful to the Aspen Center for Physics where our work was partly supported by National Science Foundation grant PHY-1066293, and by a grant from the Simons Foundation (D.M. and M.Z.). \\

\end{document}